\documentstyle[11pt,appb,epsf]{article}
\begin{document}
\date{November 15, 1996}
\title{MSX - A MONTE-CARLO CODE FOR NEUTRON EFFICIENCY CALCULATIONS FOR LARGE 
VOLUME Gd-LOADED LIQUID SCINTILLATION DETECTORS
\thanks{Presented at the XXXI Zakopane School 
of Physics, Zakopane, Poland, September 3--11, 1996.}}
\author{\underline{A.~Trzci\'nski} and B.~Zwiegli\'nski
\address{Soltan Institute for Nuclear Studies, 00-681 Warsaw, Ho\.za 69, Poland\\
{\sc for the ALADIN Collaboration:}\\GSI-Darmstadt, University of Frankfurt, 
University of Catania, University of Milano, Michigan State University, 
FZ Rossendorf, SINS Warsaw}}
\maketitle
\begin{abstract}
Some properties of the code newly developed to simulate the neutron detection 
process in a NMM are briefly described.
\end{abstract}
\PACS{29.40.Mc, 29.85.+c}
\section{Introduction}
Recent ALADIN Collaboration experiments [1] presented an extensive evidence 
that the hot projectile prefragments created in the peripheral to semicentral 
relativistic heavy-ion collisions are thermally equilibrated. Temperature 
{\it (T)} and excitation energy {\it (E$_{x}$)} are the essential variables 
specifying their thermodynamic properties. The dependence of {\it T} on 
{\it E$_{x}$} (caloric curve) revealed a plateau [2] interpreted as
indication of the liquid-gas phase transition in finite nuclear systems.
Therefore, a progress in measurement of {\it E$_{x}$} and {\it T} is a necessary 
prerequisite for the further comprehensive study of the phenomenon.

The predictions of Friedman [3] indicate that an excitation energy of a 
prefragment, whose mass and charge are generally unknown, can be determined 
by means of the multiplicities of the emitted neutrons and light charged 
particles {\it (Z$\leq$}2{\it )}. We intend to apply this method to the 
{\it target spectators} by employing a large-volume Gd-loaded liquid 
scintillation detector (a neutron multiplicity meter - NMM) to count neutrons 
originating from their decay. A study of feasibility and limitations of the 
Friedman's method in the ALADIN environment asked for a program giving 
a more comprehensive description of the neutron detection process in a NMM 
than offered by DENIS [4], presently used in several laboratories [5]. 
These considerations prompted us to develop MSX, a Monte-Carlo code 
containing several improvements in comparison with its predecessor.
\section{Neutron detection process in a NMM}
The detection process in a NMM [5] consists of the following stages:\\
{\it i)} Neutron slowing down due to the multiple scattering on 
the hydrogen and carbon nuclei of the scintillator. Ultimately, the neutron 
comes into thermal equilibrium with the medium and diffuses within the 
scintillator volume.

\begin{figure}[hhh]
\epsfxsize=11cm
\epsfbox{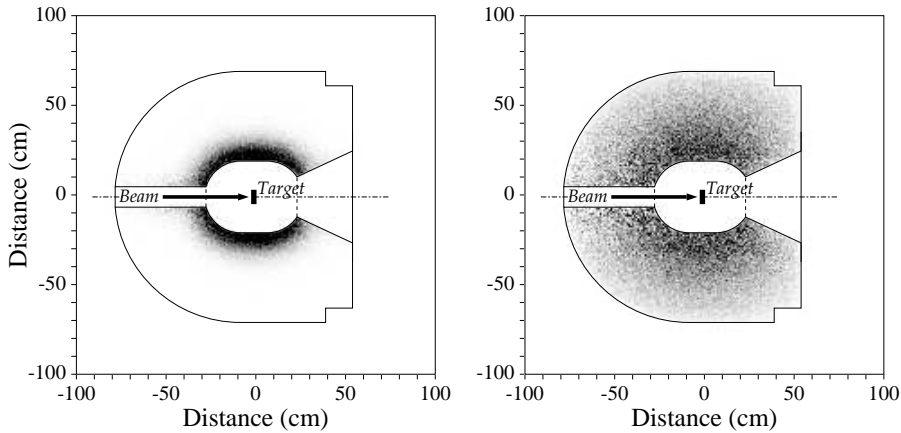}
\caption{Distribution of capture events within the volume of the tank for 
1~MeV (left) and 30~MeV (right) neutrons.}
\end{figure}
\noindent{\it ii)} Radiative capture following one of the three possibilities:
in $^{1}$H with the emission of a single 2.224~MeV $\gamma$-ray, in $^{155}$Gd 
or $^{157}$Gd accompanied by the emission of a cascade of $\gamma$-rays 
with the summed energies of 8.536 and 7.937~MeV, respectively.\\
{\it iii)} Gamma-ray thermalization due to the consecutive Compton scattering
from the electrons of H and C atoms of the scintillator.\\
{\it iv)} Electron stopping in the medium. Part of the electron's energy loss,
determined by the scintillation efficiency, is converted into visible light.\\
{\it v)} Transport of light to the photocathodes of photomultipliers (PMs)
through multiple reflections on the diffusely reflecting coating of
the interior of the tank.  

The above stage {\it (i)} involves two vastly different time scales. 
The dissipation of neutron's total initial kinetic energy occurs within a 
nanosecond time scale, giving the initial prompt part of the
scintillation signal. The signal coming from the $\gamma$-rays 
follows with a random delay of up to 50~$\mu$s, the time scale determined 
by the probability of capture, terminating the diffusion process.
Therefore essential for the operation of a NMM is the stretching of
the capture probability over tens of microseconds which allows to count 
one-by-one over hundred neutrons emitted in the interaction. 
\section{Novelties in MSX}
\noindent$\bullet$ {\it Throughout the program}\\
A. Algorithm optimized to maximize the calculational speed.\\
B. Memory utilization optimized with the aid of pointers allows to track 
simultaneously (practically) an arbitrary number of particles.\\
$\bullet$ {\it Stage (i)}\\
A. Use of the cross section files supplied by the international 
Nuclear Data Centers. They contain periodically updated data for many
reactions on H, C, and Gd (23 total and 5 differential cross sections used
by us).\\
B. Exact treatment of {\it d}$\sigma$/{\it d}$\Omega$ for: 
$^{1}$H({\it n},{\it n})$^{1}$H, $^{12}$C({\it n},{\it n})$^{12}$C, 
$^{12}$C({\it n},{\it n'})$^{12}$C (first, second and third excited state). 
This permits to follow the main tendencies of the angular distributions - 
rapid changes across the groups of prominent resonances in the 
{\it n}~+~$^{12}$C interaction at low energies and increasing forward 
peaking as the energy exceeds the resonance region.\\
C. Multiple neutron emission in the $^{12}$C({\it n},2{\it n})$^{11}$C 
reaction taken into account. It permits to determine the number of misclassified 
events (two neutrons detected instead of the incident one).\\
$\bullet$ {\it Stage (ii)}\\
A. Radiative capture in the resonance region for the
$^{155}$Gd({\it n},$\gamma$)$^{156}$Gd and
$^{157}$Gd({\it n},$\gamma$)$^{158}$Gd reactions included.\\
B. $\gamma$-ray cascades following neutron capture taken into account by 
combining the available experimental data with the calculated continuous part
of the spectrum. Calculations were done on base of the Hauser-Feshbach
statistical model for the decay of the {\it J}$^{\pi}$= 2$^{-}$
capturing state. Besides mean multiplicities, defining mean capture
$\gamma$-ray energies, also fluctuations around the mean taken into account.\\
$\bullet$ {\it Stage (v)}\\
A. Simulation of light transport to the PM photocathodes incorporated into 
the program. It permits to determine the effect of electronic thresholds 
on the detection efficiency. Moreover, the distribution of photon arrival 
times determines the PM current pulse shape, thereby defining the 
double-pulse resolution.\\
B. Calorimetric mode (in the stage of implementation). It will permit to 
simulate the dependence of the prompt peak on the energy deposited
in the scintillator by the incident neutrons.
\section{Representative results for the selected geometry}
The geometry of a NMM destinated to count neutrons from the target residue 
in coincidence with the charged products of the projectile-target interaction 
is shown in Fig.~1. The escape cone with 50$^{\circ}$ opening angle enables 
undisturbed detection of the intermediate mass fragments from the
\begin{figure}[h]
\hspace{2.7cm}
\epsfxsize=6cm
\epsfbox{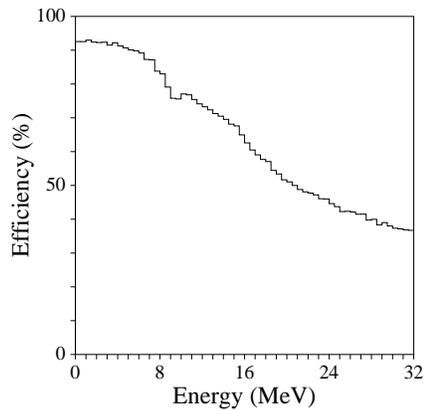}
\caption{Energy dependence of the neutron detection efficiency in the energy 
range 0.5-32~MeV.}
\end{figure}
projectile-like source by means of the ALADIN heavy fragment detectors.
Light charged particles arising from the intermediate-velocity source
are registered with the 92-element Si-CsI(Tl) array [1]. These are used
to tag the impact parameter involved in the collision. In the same time
neutrons from the latter two sources are prevented from interacting with
the scintillator and the tank walls. The results of MSX simulations for
the indicated geometry are presented in Figs.~1 and~2.


\begin{thebibliography}{99}
\bibitem{}
A.~Sch\"{u}ttauf {\it et al.}, {\it Nucl.~Phys.} {\bf A607}, 457 (1996).
\bibitem{}
J.~Pochodzalla {\it et al.}, {\it Phys.~Rev.~Lett.} {\bf 75}, 1040 (1995).
\bibitem{}
W.A.~Friedman, {\it Phys.~Lett.} {\bf B242}, 309 (1990).
\bibitem{}
J.~Poitou, H.~Nifenecker, C.~Signarbieux, {\it Report} {\bf CEA-N-1282} (1970);
J.~Poitou, C.~Signarbieux, {\it Nucl.~Instr.~and~Meth.} {\bf 144}, 113 (1974).
\bibitem{}
J.~Galin, U.~Jahnke, {\it J.~Phys.~G:~Nucl.~Part.~Phys.} {\bf 20}, 1105 (1994).
\end{thebibliography}
\end{document}